# THE LOSCHMIDT PARADOX

# ON BOLTZMANN´S H-THEOREM:

# A RESOLUTION

**Evangelos Chaliasos**

365 Thebes Street

GR-12241 Aegaleo

Athens, Greece

*Abstract*

A resolution of the Loschmidt paradox on Boltzmann´s H-theorem is proposed, based on some new concepts concerning motion backwards in time. The key to this resolution, taken from our interpretation of "motion backwards in time", is the associated notion of *negative mass,* attributed to particles moving backwards in time, which we call *retro-particles,* or simply *retrons*.

## 1. Introduction

It is well known the Boltzmann H-theorem. According to this theorem, in a gas with velocity distribution f(**v**, t), the H-function, that is the quantity

$$H = \int_{-\infty}^{+\infty}\int_{-\infty}^{+\infty}\int_{-\infty}^{+\infty} d^3v f(\vec{v},t) \ln f(\vec{v},t), \qquad (1)$$

decreases with time, that is
$$dH/dt \leq 0. \qquad (2)$$

One of the early classical objections to this theorem was the reversibility objection raised by Loschmidt [1, 2, 3]. According to this objection the H-theorem must be incorrect because of the well known invariance of the laws of Mechanics under the time reversal t → t′ = -t. Thus, it must be incorrect because the reversibility at the microscopic level, on which it is based, cannot lead to macroscopic irreversibility, which is expressed by the relation (2).

To be more clear, because the individual collisions in the gas are temporally reversible, if we reverse the motions of the particles constituting the gas at the same time, that is if we simply reverse the direction of time, by using the time t′ = -t, the system will pass through the same states, but in the opposite order, and the new H-function, say H′, should increase, that is it should
$$dH'/dt' \geq 0. \qquad (3)$$

But, as it is evident, the velocities do not change[*], that is **v** → **v**′ = **v**, while t → t′ = -t in the definition (1), so that H → H′ with

$$H' = \int_{-\infty}^{+\infty}\int_{-\infty}^{+\infty}\int_{-\infty}^{+\infty} d^3v f(\vec{v},t') \ln f(\vec{v},t'), \qquad (4)$$

and the H-theorem gives again

---

[*] If the motion backwards in time is described by the same world line with the motion forward in time, but followed in the opposite direction of course, we would have at the same time d**x** → d**x**′ = -d**x** for the spatial coordinates, besides dt → dt′ = -dt, so that **v** → **v**′ = **v** by the definition of velocity.

$dH'/dt' \leq 0$. (5)

It is the purpose of this paper to present a resolution of the above paradox.

### 2. Motion backwards in time and negative mass

We have from Mechanics for the action S

$$S = \int_{t_A}^{t_B} L\,dt \quad \text{for motion forward in time } (t_A < t_B),$$ (6)

where L is the Lagrangian. But we have to take

$$S = \int_{t_B}^{t_A} L'\,dt \quad \text{for motion backwards in time } (t_B > t_A),$$ (7)

where L´ is the new Lagrangian. By a comparison of eqns. (6) and (7), we get
$L' = -L$, (8)

if the motion backwards in time is described by the same world line with the motion forward in time (but followed in the opposite direction of course).

But, for a free particle,

$$L = \frac{1}{2}mv^2, \quad \text{with } m > 0,$$ (9)

because, if m were negative, S could not have a minimum, required by the principle of least action, since then we could make S as small (negative) as we wanted, by choosing v sufficiently large. Thus also, because of eqn. (8), we must take

$$L' = \frac{1}{2}m'v'^2, \quad \text{with } m' < 0.$$ (10)

And, since, as we have seen in (*),
$v' = v$, (11)

we finally get, comparing eqn. (9) with eqn. (10) and because of eqn (8),
$m' = -m$, (12)

the prime attributed to the so-called *retro-particles* or *retrons* (i.e . particles moving *backwards* in time), in contrast to normal particles, which we denote without prime (particles moving forward in time):

$$m > 0 \quad \text{normal}; \quad m' < 0 \quad \text{retro}. \tag{13}$$

The momentum is defined in the usual way in both cases (normal & retro), that is

$$\vec{p} = m\vec{v} \tag{14}$$

for normal particles, and

$$\vec{p}' = m'\vec{v}' \tag{15}$$

for retroparticles. The basic feature of these formulae is that they give a momentum of the same direction with the velocity for normal particles, but a momentum and a velocity of opposite directions for retro-particles. If the world line of a retroparticle is the same (but of the opposite direction) as compared to the world line of a corresponding normal particle, then obviously **v**′ = **v** (see (*)) and, since m′ = -m, formulae (14) and (15) give opposite momenta, because they become respectively

$$\vec{p} = m\vec{v} \tag{16}$$

and

$$\vec{p}' = -m\vec{v}. \tag{17}$$

### 3. Resolution of Loschmidt´s paradox

We have for normal particles

$$H_{normal} = \int_{-\infty}^{+\infty}\int_{-\infty}^{+\infty}\int_{-\infty}^{+\infty} d^3 v f(\vec{v},t)\ln f(\vec{v},t). \qquad (18)$$

As we saw, for normal particles the H-theorem

$$dH_{normal}/dt \leq 0 \qquad (19)$$

holds. But, because of the underlying canonical formalism, we have to introduce the momenta rather, instead of the velocities. Then, because of formula (14), or (16), the definition (18) can be stated as

$$H_{normal} = \int_{-\infty}^{+\infty}\int_{-\infty}^{+\infty}\int_{-\infty}^{+\infty} \frac{d^3 p}{m} f\left(\frac{\vec{p}}{m},t\right)\ln f\left(\frac{\vec{p}}{m},t\right). \qquad (20)$$

If we now consider the case of retroparticles, we have to take, because of eqn. (15),

$$H_{retro} = \int_{-\infty}^{+\infty}\int_{-\infty}^{+\infty}\int_{-\infty}^{+\infty} \frac{d^3 p'}{m'} f\left(\frac{\vec{p}'}{m'},t\right)\ln f\left(\frac{\vec{p}'}{m'},t\right) \qquad (21)$$

rather than eqn. (20). But now, because of eqns. (17) & (16), besides eqn. (12), eqn. (21) has to be written

$$H_{retro} = \int_{p_x=+\infty}^{-\infty}\int_{p_y=+\infty}^{-\infty}\int_{p_z=+\infty}^{-\infty} \frac{d^3(-p)}{-m} f\left(\frac{-\vec{p}}{-m},t\right)\ln f\left(\frac{-\vec{p}}{-m},t\right) \qquad (22)$$

rather than eqn. (20). Thus, because of eqn. (16), we have to write, using now the velocities,

$$H_{retro} = \int_{v_x=+\infty}^{-\infty}\int_{v_y=+\infty}^{-\infty}\int_{v_z=+\infty}^{-\infty} d^3 v f(\vec{v},t)\ln f(\vec{v},t). \qquad (23)$$

Interchanging the limits of integration, eqn. (23) becomes

$$H_{retro} = -\int_{v_x=-\infty}^{+\infty}\int_{v_y=-\infty}^{+\infty}\int_{v_z=-\infty}^{+\infty} d^3 v f(\vec{v},t)\ln f(\vec{v},t), \qquad (24)$$

so that, after a comparison of eqns. (24) and (18), finally

$$H_{retro} = -H_{normal}. \qquad (25)$$

It therefore follows from eqn. (19), because of eqn. (25), that

$$dH_{retro}/dt \geq 0, \qquad (26)$$

a relation which could be called the *anti H-theorem*.

Now we reverse the time and use the time t´ = -t. We also use a prime in the new H-functions in order to distinguish them from the old ones. It is obvious that the normal particles with respect to the time t will be retroparticles with respect to the time t´ on the one hand, and on the other hand conversely the retroparticles in the time t will be normal particles in the time t´. Thus we have the relations

$$H'_{retro} = H_{normal} \qquad (27)$$

and

$$H'_{normal} = H_{retro} . \qquad (28)$$

Therefore, concerning normal particles in the time t, combination of eqns. (19) (H-theorem) and (27) gives

$$\frac{dH'_{retro}}{dt'} = \frac{dH_{normal}}{d(-t)} \geq 0 , \qquad (29)$$

in accordance with Loschmidt´s relation (3). Boltzmann´s relation (5) corresponds to normal particles in the time t´. Combination of eqns. (26) (anti H-theorem) and (28) gives in this case

$$\frac{dH'_{normal}}{dt'} = \frac{dH_{retro}}{d(-t)} \leq 0 . \qquad (30)$$

Thus for motion backwards in time (retroparticles) we have to refer to the anti H-theorem rather than to the H-theorem (normal particles). In this way, the anti H-theorem gives eqn. (30), despite the fact that the H-theorem gives eqn. (29). This latter theorem refers *only* to normal particles, while *for retrons, i.e. for motion backwards in time, the anti-theorem holds*. Thus the resolution of Loschmidt´s paradox has been achieved.

**Appendix: relation of H with entropy**

If we make use of the probability p instead of the distribution function f, we have from the definition (1) the relation

$$H = \int d^3v \frac{N}{V} p(\vec{v},t) \ln\left[\frac{N}{V} p(\vec{v},t)\right], \tag{31}$$

where N is the total number of particles and V is the volume occupied by them. Performing the integration, we obtain

$$H = \frac{N}{V} \ln \frac{N}{V} + \frac{N}{V}(-S_1)\frac{1}{k}, \tag{32}$$

where $S_1$ denotes the entropy contributed by one particle and k is the Boltzmann constant. Thus finally

$$S = -kVH + const. \tag{33}$$